\documentclass[copyright,creativecommons]{eptcs}
\providecommand{\event}{ACL2 2013} 
\usepackage{breakurl}             

\usepackage{amsmath}
\usepackage{amssymb}

\usepackage{xy}
\xyoption{matrix}
\xyoption{frame}
\xyoption{arrow}
\xyoption{arc}

\usepackage{ifpdf}
\ifpdf
\else
\PackageWarningNoLine{Qcircuit}{Qcircuit is loading in Postscript mode.  The Xy-pic options ps and dvips will be loaded.  If you wish to use other Postscript drivers for Xy-pic, you must modify the code in Qcircuit.tex}
\xyoption{ps}
\xyoption{dvips}
\fi

\entrymodifiers={!C\entrybox}

\newcommand{\ket}[1]{{\left\vert{#1}\right\rangle}}
\newcommand{\qw}[1][-1]{\ar @{-} [0,#1]}
\newcommand{\qwx}[1][-1]{\ar @{-} [#1,0]}
\newcommand{\cw}[1][-1]{\ar @{=} [0,#1]}
\newcommand{\cwx}[1][-1]{\ar @{=} [#1,0]}
\newcommand{\gate}[1]{*+<.6em>{#1} \POS ="i","i"+UR;"i"+UL **\dir{-};"i"+DL **\dir{-};"i"+DR **\dir{-};"i"+UR **\dir{-},"i" \qw}
\newcommand{\meter}{*=<1.8em,1.4em>{\xy ="j","j"-<.778em,.322em>;{"j"+<.778em,-.322em> \ellipse ur,_{}},"j"-<0em,.4em>;p+<.5em,.9em> **\dir{-},"j"+<2.2em,2.2em>*{},"j"-<2.2em,2.2em>*{} \endxy} \POS ="i","i"+UR;"i"+UL **\dir{-};"i"+DL **\dir{-};"i"+DR **\dir{-};"i"+UR **\dir{-},"i" \qw}

\newcommand{\control}{*!<0em,.025em>-=-<.2em>{\bullet}}

\newcommand{\ctrl}[1]{\control \qwx[#1] \qw}

\newcommand{\targ}{*+<.02em,.02em>{\xy ="i","i"-<.39em,0em>;"i"+<.39em,0em> **\dir{-}, "i"-<0em,.39em>;"i"+<0em,.39em> **\dir{-},"i"*\xycircle<.4em>{} \endxy} \qw}

\newcommand{\gategroup}[6]{\POS"#1,#2"."#3,#2"."#1,#4"."#3,#4"!C*+<#5>\frm{#6}}

\newcommand{\rstick}[1]{*!L!<-.5em,0em>=<0em>{#1}}
\newcommand{\lstick}[1]{*!R!<.5em,0em>=<0em>{#1}}

\newcommand{\Qcircuit}{\xymatrix @*=<0em>}

\usepackage{listings}
\begin{document}

\title{An Interpreter for Quantum Circuits}
\def\titlerunning{An Interpreter for Quantum Circuits}
\def\authorrunning{L. Helms \& R. Gamboa}

\author{Lucas Helms
\institute{University of Wyoming \\
Laramie, WY, USA}
\email{lhelms3@uwyo.edu}
\and
Ruben Gamboa
\institute{University of Wyoming \\
Laramie, WY, USA}
\email{ruben@uwyo.edu}
}

\maketitle
\begin{abstract}
    This paper describes an ACL2 interpreter for ``netlists'' describing quantum circuits.
    Several quantum gates are implemented, including the Hadamard gate H, which rotates
    vectors by $45^\circ$, necessitating the use of irrational numbers, at least at the
    logical level.  Quantum measurement presents an especially difficult challenge, because
    it requires precise comparisons of irrational numbers and the use of random numbers.  
    This paper does not address computation with irrational numbers or the generation of 
    random numbers, although future work includes the development of pseudo-random 
    generators for ACL2.
\end{abstract}




\section{Background}

Quantum computing is a rich field, blending physics, linear algebra, randomness, and computation.
It can be approached from many different angles, and some treatments delve considerably into the
underlying quantum physics.  However, the field is sufficiently advanced that high-level concepts
permit a discussion of quantum computing in a computer science context, i.e., algorithms can be
specified as circuits with quantum analogues of bits, wires, and 
gates~\cite{Vazirani:quantum-notes,YaMa:quantum-computing,McMahon:quantum-computing}.
In this paper, we use these high-level concepts to model quantum circuits in ACL2~\cite{KM:ACL2-industrial}.

The most fundamental concept is that of a qubit, the quantum analogue of a bit of information.
A qubit can have the values $\ket{0}$ or $\ket{1}$.  These are the ``major'' possibilities, 
corresponding to \texttt{T} and \texttt{NIL}, for example.  But qubits can also be in a quantum
superposition, which corresponds to a linear combination of states, e.g., $\alpha\ket{0} + \beta\ket{1}$,
where $\alpha$ and $\beta$ are \emph{complex} (not just real) numbers.  There is a useful restriction
that $||\alpha||^2 + ||\beta||^2 = 1$, where $||\cdot||$ denotes the complex norm.

Thinking in terms of linear algebra, $\ket{0}$ and $\ket{1}$ are a set of basis vectors (cf.~$x$ 
and $y$ unit vectors) and a qubit can be in the space spanned by these vectors, where the scalar 
field is taken from the complex numbers (subject to the norm restriction above).  Because of the
norm restrictions, the space of a qubit can be visualized as a sphere.

An important operation is that of measuring a qubit.  In general, measurements can be taken with
respect to any basis, but we will restrict ourselves to the $\ket{0}$--$\ket{1}$ basis.  The result
of a measurement is twofold.  First, the measurement will yield only one of the possible answers
$\ket{0}$ or $\ket{1}$, where $\ket{0}$ is (randomly) chosen with probability $||\alpha||^2$ and 
$\ket{1}$ with probability $||\beta||^2$.  Second, the qubit will ``collapse'' to the measured
value.  For example, suppose we measure the qubit $\frac{1}{\sqrt{3}}\ket{0} + \sqrt{\frac{2}{3}}\ket{1}$.
Then with probability of $\frac{1}{3}$, the result will be $\ket{0}$, \emph{and} the qubit will
become $\ket{0}$, so that subsequent measurements of this qubit will also be $\ket{0}$.  Similarly,
with probability $\frac{2}{3}$, the result will be $\ket{1}$ and the qubit will collapse to $\ket{1}$.

Quantum gates operate on qubits.  Unlike classical gates, quantum gates always have the same input and
output arity.  I.e., there are 1-qubit gates, 2-qubit gates, and so on, but 1-qubit gates always have
one output, 2-qubit gates always have two outputs, etc.  In our previous discussion, we introduced the
first quantum gate, namely a measurement gate.  Other unary gates include the $X$ gate, which is analogous
to logical negation, in that it maps $\ket{0}$ to $\ket{1}$ and vice versa.  More generally, $X$ maps
$\alpha\ket{0} + \beta\ket{1}$ to  $\beta\ket{0} + \alpha\ket{1}$.

The $Z$ gate has no classical analogue.  It flips the sign of the $\ket{1}$ component, so that 
$\alpha\ket{0} + \beta\ket{1}$ is mapped to  $\alpha\ket{0} - \beta\ket{1}$.

The $X$ and $Z$ gates merely rearrange the coordinates of a qubit in the $\ket{0}$--$\ket{1}$ space.
The Hadamard or $H$ gate, on the other hand, is best described as a $45^\circ$ rotation.  It maps
$\alpha\ket{0} + \beta\ket{1}$ to $\alpha\frac{\ket{0}+\ket{1}}{\sqrt{2}} + \beta\frac{\ket{0}-\ket{1}}{\sqrt{2}}
= \frac{\alpha+\beta}{\sqrt{2}}\ket{0} + \frac{\alpha-\beta}{\sqrt{2}}\ket{1}$.  We will soon see
that far from being a curiosity, the $H$ gate plays a crucial role in quantum circuits.
Because this gate is defined in terms of the irrational $\sqrt{2}$, our logical model of quantum circuits is
built in ACL2(r)~\cite{GaKa:acl2r}.

Naturally, circuits can be built using more than one qubit.  When more than one qubit is present,
the state of the qubits is represented using basis vectors in a higher-dimensional space.  For
instance, if two qubits are present, then the basis vectors are $\ket{00}$, $\ket{01}$,
$\ket{10}$ and $\ket{11}$.  The state of the qubits is represented by $\alpha\ket{00} + \beta\ket{01}
+ \gamma\ket{10} + \delta\ket{11}$, subject to $||\alpha||^2 + ||\beta||^2 + ||\gamma||^2 + ||\delta||^2 = 1$ as before.

While at first the combination of qubits looks very much like the classical combination of bits,
the quantum world offers some unique twists.  This is best illustrated by considering measurement.
Suppose two qubits are in the state $\frac{1}{2}\ket{00} + \frac{1}{2}\ket{01}
+ \frac{1}{2}\ket{10} + \frac{1}{2}\ket{11}$ and the first qubit is measured.  All outcomes are
equally likely, so suppose that the result of the measurement is $\ket{1}$.  As discussed previously,
this results in a collapse of the first qubit to $\ket{1}$, so the resulting quantum state\footnote{
Notice that the coefficients are ``renormalized'' after the measurement, so that the resulting state
satisfies the requirements that the sum of the square of the norms equals one.} is 
$\frac{1}{\sqrt{2}}\ket{10} + \frac{1}{\sqrt{2}}\ket{11}$. Notice that any subsequent measurements
of the first qubit will necessarily result in $\ket{1}$, as expected.

But consider, instead, the quantum state $\frac{1}{\sqrt{2}}\ket{01} + \frac{1}{\sqrt{2}}\ket{10}$.
Again, suppose that the first qubit is measured and that the result is a $\ket{1}$.  As before,
the quantum state collapses so that the first qubit is forced to be $\ket{1}$.  Only this time,
the collapse is such that the \emph{second} qubit also collapses, in this case to $\ket{0}$.
I.e., the quantum state (after normalizing the coefficients) collapses to $\ket{10}$, so both
qubits are in a known state.  This phenomenon is known as \emph{entanglement}, and it is the
reason that quantum states cannot always be decomposed into substates involving fewer qubits.
For instance, quantum state $\frac{1}{\sqrt{2}}\ket{01} + \frac{1}{\sqrt{2}}\ket{10}$ cannot
be decomposed into two substates, one for each qubit.

Now that we have considered quantum states with more than one qubit, we can return to the description
of quantum gates.  The controlled not gate, or $CN$ gate, operates on two qubits, called the control
qubit and the target qubit.  This gate leaves the control qubit unchanged, but it has the effect of
a not ($X$) gate on the target qubit precisely when the control qubit is a $\ket{1}$.  It is important
to note that the $CN$ gate does this without measuring the first qubit; otherwise, it wouldn't 
leave it unchanged.  For example, suppose two qubits are in the state
$\frac{1}{\sqrt{2}}\ket{01} + \frac{1}{\sqrt{2}}\ket{10}$, and that we pass these qubits through a
$CN$ gate where the first qubit is the control qubit and the second is the target.  The resulting
state will be $\frac{1}{\sqrt{2}}\ket{01} + \frac{1}{\sqrt{2}}\ket{11}$, where the second qubit was
``negated'' when the first qubit is a $\ket{1}$.  Notice that the two
qubits are no longer entangled, although (in this particular example) the state of the second qubit
is now determined.

Of course, our discussion so far only scratches the rich field of quantum circuits.  For instance,
many more gates are possible.  In fact, it is fruitful to think of quantum states as vectors and
gates as unitary matrices, in which case the effect of a gate on a quantum state is reduced to
matrix multiplication.  There are as many different gates as there are unitary matrices!
Similarly, the process of measuring a qubit can be construed using Hermitian
matrices, and the result of a measurement is a Eigenvalue of the matrix and the resulting collapse is
to the projection of the state vector into the corresponding Eigenspace.

Nevertheless, this brief introduction to quantum circuits is quite sufficient to model and reason about 
them in ACL2.  We will describe our logical model of quantum circuits in Sect.~\ref{circuits-logic}.  
This introduces ACL2 functions that model quantum states, quantum gates, and quantum circuits.  We
will illustrate this with an ACL2 model of the quantum teleportation circuit in Sect.~\ref{teleportation},
where we prove that the teleportation circuit does in fact result in the (destructive) copy of the state
of one qubit into another one.  As observed previously, the logical model uses irrational numbers, such
as $\sqrt{2}$.  Although ACL2(r) supports reasoning over the irrationals, it does not offer a solution
for computing with such numbers.  What this means is that the logical model we developed is precisely
that, a logical model without an executable counterpart.  In Sect.~\ref{evaluator}, we show one possible 
way to resolve this by using rational approximations, so that we can at least approximate the result of
a quantum circuit.  However, the result is not entirely satisfactory, so we conclude in Sect.~\ref{conclusion}
with some ideas for improving our ACL2 model of quantum computation in the future.

\section{Quantum Circuits: Logical Model}
\label{circuits-logic}

We begin with a description of quantum states, as modeled in ACL2.  In general, a quantum state can be
modeled as $\sum_{x}{\alpha_{x}\ket{x}}$, where $x$ ranges over all possible qubit configurations, e.g.,
for a 3-qubit system, $x \in \{\ket{000}, \dots, \ket{111}\}$.  So one possible representation is
simply the corresponding vector of complex numbers $\langle \alpha_x \rangle$.  This representation has
much to recommend it, and it is often used in textbooks.  However, it proves cumbersome to use this
representation to implement, for example, the Hadamard gate.

So we chose to use a more ``algebraic'' representation, with explicit coefficients $\alpha_x$ and basis vectors 
$\ket{x}$.  For instance, we represent the state $\frac{1}{\sqrt{2}}\ket{01} + \frac{1}{\sqrt{2}}\ket{10}$
with the ACL2 list
\begin{lstlisting}
( ((/ (acl2-sqrt 2)) NIL T  )
  ((/ (acl2-sqrt 2)) T   NIL) )
\end{lstlisting}
Note that this state can also be represented as 
\begin{lstlisting}
( (0 NIL NIL)
  ((/ (acl2-sqrt 2)) NIL T  )
  ((/ (acl2-sqrt 2)) T   NIL)
  (0 T T) )
\end{lstlisting}
and even as 
\begin{lstlisting}
( (0 NIL NIL)
  ((* 1/3 (/ (acl2-sqrt 2))) NIL T  )
  ((/ (acl2-sqrt 2)) T   NIL)
  ((* 2/3 (/ (acl2-sqrt 2))) NIL T  )
  (0 T T) )
\end{lstlisting}
The only restriction is that each term in the list consists of a (possibly) complex magnitude and precisely as many
booleans as there are qubits in the system.
We use the function \texttt{sort-and-merge} to convert any of these representations into the
preferred representation, which is the second one, with one (possibly zero) entry for each basis vector.

Recall that the coefficients of any valid quantum state are restricted so that the sum of the squares of 
their complex norm must be equal to one.  The function \texttt{qustate-norm} computes the norm of a
quantum state, which is the square root of the previous sum.  This can be used to scale any non-zero
quantum state so that the coefficients have the right property, and the function \texttt{qustate-normalize}
does this.

For each gate, we define a function that computes it.  Because we intend to use gates in quantum circuits,
the input to the gates is not a single qubit, but an entire quantum state possibly comprising several qubits.
So part of the argument to the function is the (zero-based) index of the qubit or qubits that apply.  For
example, the $X$ gate is implemented with the function \texttt{(qu-X-gate s n r)}, where \texttt{s} denotes
a quantum state, \texttt{n} is the qubit on which the $X$ gate should operate, and \texttt{r} is 
unused\footnote{It's only used by the measurement gates, but we included it in all gates for uniformity.}.
Similarly, the $CN$ gate is implemented by the function \texttt{(qu-CN-gate s c n r)}, where \texttt{c}
is the control qubit and \texttt{n} the target qubit.  All of these functions return the quantum state
resulting from the execution of the appropriate gate.

A qubit can be measured using the $M$ gate, which is implemented by the function \texttt{(qu-M-gate s n r)}.
Here, the argument \texttt{r} is
necessary.  It should be a real number from $0$ to $1$, used to index into a cumulative distribution function.  
Suppose
that the qubit is in state $\alpha\ket{0} + \beta\ket{1}$.  Then the measurement will result in
$\ket{0}$ if the argument \texttt{r} is less than $||\alpha||^2$, and in $\ket{1}$ otherwise.
As with other gates, this returns the quantum state resulting 
from the measurement, as opposed to the actual measurement.  Naturally, the resulting state is such that
the measured qubit can have only one value.  This value can be extracted using the function 
\texttt{(get-deterministic-qubit s n)}.

A (possibly flawed) design decision we made is that the result of the gate operations is not necessarily
a canonical quantum state.  If such a state is required, the caller must call \texttt{qustate-normalize}
explicitly.

\begin{figure}
\begin{align*}
\Qcircuit @C=.7em @R=.4em @! {
  \lstick{\ket{0}} & \gate{H} & \ctrl{1} & \rstick{\frac{1}{\sqrt{2}}\ket{00} + \frac{1}{\sqrt{2}}\ket{11}} \qw\\
  \lstick{\ket{0}} & \qw & \targ & \qw
 }
\end{align*}
\caption{Quantum Circuit}
\label{quantum-circuit}
\end{figure}
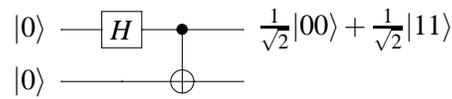

Quantum circuits are built by successively applying individual quantum gates to selected qubits.  For
example, Fig.~\ref{quantum-circuit} shows a simple quantum circuit involving two qubits.  As the figure
shows, the qubits are both initially $\ket{0}$.  Then the first qubit is passed through an $H$ gate,
and then the two qubits are passed through a $CN$ gate, where the first qubit is the control qubit
and the second the target.  The resulting state is an entangled state, so it cannot be described by
giving the values of each individual qubit separately.  Notice how the Hadamard gate is key to
entangling qubits.  This is one of the reasons why it is fundamental in quantum circuits.

We represent these circuits in ACL2 using a list of gate applications.  Each gate application is a
list consisting of the gate type and the qubit (or qubits) to which the gate applies.  For example,
the circuit in Fig.~\ref{quantum-circuit} is represented with the following list
\begin{lstlisting}
( (H 0)
  (CN 0 1) )
\end{lstlisting}
Unlike classical gates, quantum gates always have the same number of input and output qubits.  So it
is natural to linearize quantum circuits into this representation.

The function \texttt{(qustate-circuit circuit qstate rs)} implements a quantum circuit.  The argument
\texttt{circuit} is a list describing the circuit to be evaluated, \texttt{qstate} is the starting
quantum state, and \texttt{rs} is a list of random numbers in the range $[0,1]$.  This list should have
at least as many random numbers as the circuit has $M$ gates.

Essentially, the function \texttt{qustate-circuit} simply applies each gate in the circuit to the
quantum state.  The only complication is that the state must be normalized between successive gates,
since the output of the gates may be unnormalized.  The main loop is implemented in the following function:
\begin{lstlisting}
(defun qstate-circuit-aux (circuit qstate rs)
  (if (endp circuit)
      qstate
    (let ((op (caar circuit))
          (args (cdar circuit)))
      (qstate-circuit-aux 
       (cdr circuit)
       (qustate-normalize
         (case op
            (X  (qu-X-gate qstate (car args) 0))
            (Z  (qu-Z-gate qstate (car args) 0))
            (CN (qu-CN-gate qstate (car args) 
                                   (cadr args) 
                                   0))
            (H  (qu-H-gate qstate (car args) 0))
            (I  (qu-I-gate qstate (car args) 0))
            (M  (qu-M-gate qstate (car args) 
                                  (car rs)))))
       (if (equal op 'M)
           (cdr rs)
         rs)))))
\end{lstlisting}
Notice that only the $M$ gate consumes random numbers from the source.
The function \texttt{qstate-circuit} simply normalizes the initial
quantum state and then calls \texttt{qstate-circuit-aux}.

\section{A Quantum Teleportation Circuit}
\label{teleportation}

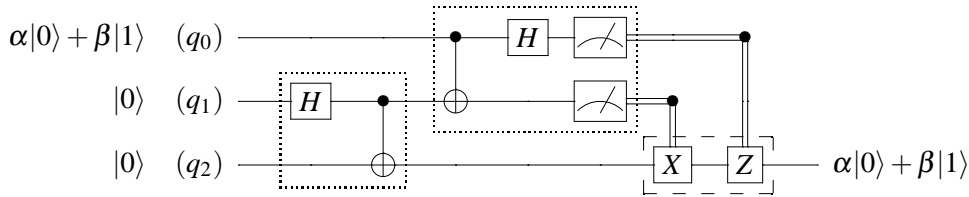
\begin{figure*}
\begin{align*}
\Qcircuit @C=.7em @R=.4em @! {
  \lstick{\alpha\ket{0}+\beta\ket{1} \quad (q_0)} & \qw & \qw & \ctrl{1}&
    \gate{H} & \meter & \cw &  \control \cw & \\
  \lstick{\ket{0} \quad (q_1)} & \gate{H} & \ctrl{1} & \targ & \qw &
    \meter & \control \cw & \cwx & \\
  \lstick{\ket{0} \quad (q_2)} & \qw & \targ & \qw &
    \qw & \qw & \gate{X} \cwx & \gate{Z} \cwx &
    \rstick{\alpha\ket{0}+\beta\ket{1}} \qw
	\gategroup{2}{2}{3}{3}{.7em}{.}
	\gategroup{1}{4}{2}{6}{.7em}{.}
	\gategroup{3}{7}{3}{8}{.7em}{--}
 }
\end{align*}
\caption{Quantum Teleportation Circuit}
\label{teleportation-circuit}
\end{figure*}

We are now ready to reason about a quantum circuit in ACL2.  Fig.~\ref{teleportation-circuit}
illustrates the quantum teleportation circuit.  The idea behind this circuit is that
it enables the state of a qubit to be transported into a different qubit without causing the
qubit to collapse into $\ket{0}$ or $\ket{1}$.

As is traditional in the literature, we will use the fictitious quantum physicists Alice
and Bob to describe the algorithm.  The algorithm consists of three distinct parts, and
these are shown as groups in Fig.~\ref{teleportation-circuit}.  In the first part, Alice
and Bob entangle qubits $q_1$ and $q_2$. In a ``practical'' application 
of this algorithm, Alice and Bob would entangle these qubits a priori, and then Bob could take his
qubit (carefully shielded so as to prevent accidental measurement) to a separate lab,
possibly far away from Alice.
In the second part, Alice performs a series of gates and measurements on qubits $q_0$ and $q_1$,
effectively destroying both.  But this also has the side-effect of transforming Bob's
qubit into a form that is nearly identical to the initial value of qubit $q_0$.  What remains
is for Alice to tell Bob the results of her measurements of qubits $q_0$ and $q_1$ (after the
other gate transformations), so that Bob can perform the appropriate transformations to
make qubit $q_2$ end up in \emph{exactly} the same state that qubit $q_0$ started in.  The effect
is that the state of qubit $q_0$ is ``teleported'' into qubit $q_2$, even though the two qubits
could be very widely separated.  Note that since this algorithm requires that Alice tell
Bob about the result of her measurements, this algorithm does not result in faster-than-light
communication.

We broke this algorithms into two functions, corresponding to the actions taken by Alice
and Bob.  We chose, somewhat arbitrarily, to merge the first two steps of the algorithm
into Alice's function.  This is seen in Fig.~\ref{teleportation-circuit}, where Alice's
actions are enclosed in a dotted box, and Bob's actions in a dashed box.  The three boxes
correspond to the three phases of the algorithm, as described above.

Alice's portion of the work is captured by the following function:
\begin{lstlisting}
(defun quantum-teleportation-alice (alpha beta r1 r2)
  (qstate-circuit 
   '( (H 1)
      (CN 1 2)
      (CN 0 1)
      (H 0)
      (M 0)
      (M 1) )
   (qstate-tensor-product (make-qubit alpha beta)
                          (zero-qstate 2))
   (list r1 r2)))
\end{lstlisting}
The tensor product simply sets up the initial state, with qubit $q_0$ in
state $\alpha\ket{0} + \beta\ket{1}$ and qubits $q_1$ and $q_2$ in the zero
state $\ket{0}$.  The arguments \texttt{r1} and \texttt{r2} are random
numbers in $[0,1]$, and they will be used to take the two measurements
in the circuit.

The proof that the teleportation algorithm works correctly proceeds by
successively unfolding the recursion of Alice's implicit call to 
\texttt{qstate-circuit-aux}.  That is, it consists of eight main steps.
The first step simply establishes that the initial state is already
normalized, so that the call to \texttt{qstate-circuit} simply passes
through to \texttt{qstate-circuit-aux}.  The remaining seven steps
correspond to a symbolic execution of each gate on the intermediate
quantum state.  The final result is summarized below:
\begin{lstlisting}%[captionpos=b,caption=The Result of Alice's Part of the Teleportation Algorithm, label=alice, float=*]
(defthm quantum-teleportation-alice-value
  (implies (and (acl2-numberp alpha)
                (acl2-numberp beta)
                (equal (+ (* alpha (conjugate alpha)) 
                          (* beta  (conjugate beta)))
                       1))
           (equal (quantum-teleportation-alice alpha beta r1 r2)
                  (if (< r1 1/2)
                      (if (< r2 1/2)
                          (list (list alpha NIL NIL NIL)
                                (list beta  NIL NIL T  )
                                (list 0     NIL T   NIL)
                                (list 0     NIL T   T  )
                                (list 0     T   NIL NIL)
                                (list 0     T   NIL T  )
                                (list 0     T   T   NIL)
                                (list 0     T   T   T  ))
                        (list (list 0     NIL NIL NIL)
                              (list 0     NIL NIL T  )
                              (list beta  NIL T   NIL)
                              (list alpha NIL T   T  )
                              (list 0     T   NIL NIL)
                              (list 0     T   NIL T  )
                              (list 0     T   T   NIL)
                              (list 0     T   T   T  )))
                    (if (< r2 1/2)
                        ...
                      ...))))
  :hints ...
  )
\end{lstlisting}
As this theorem clearly shows, the result of Alice's efforts is to leave the
qubits in one of four possible states, corresponding to the possible
results of the measurements of qubits $q_0$ and $q_1$.  

Of course, these measurements are dictated by the value of the random numbers
\texttt{r1} and \texttt{r2}.  However, these random numbers should be inherently
unknown to Alice.  I.e., Alice cannot see the dice with which nature is playing.
This does not present a problem, however, because Alice can see the result of
the measurement.  For instance, when the measurements
are $\ket{0}$ and $\ket{0}$, qubit $q_2$ is in state $\alpha\ket{0} + \beta\ket{1}$,
whereas when the measured values are $\ket{0}$ and $\ket{1}$, qubit $q_2$ is in state
$\beta\ket{0} + \alpha\ket{1}$.

To complete the circuit, Bob now needs to transform qubit $q_2$ into the state
$\alpha\ket{0} + \beta\ket{1}$.  He can do this by performing some transformations
based on the results of Alice's measurement of qubits $q_0$ and $q_1$, which she transmits
using classical means.  Specifically, Bob will perform an $X$ gate on qubit $q_2$ if
qubit $q_1$ was measured as $\ket{1}$.  In addition, Bob will perform a $Z$ gate on qubit
$q_2$ if qubit $q_0$ was a $\ket{1}$.  This is codified in the following function:
\begin{lstlisting}
(defun quantum-teleportation-bob (qstate q0 q1)
  (cond ((and (not q0) (not q1))
         qstate)
        ((and (not q0) q1)
         (qstate-circuit '( (X 2) ) qstate nil))
        ((and q0 (not q1))
         (qstate-circuit '( (Z 2) ) qstate nil))
        ((and q0 q1)
         (qstate-circuit '( (X 2) (Z 2) ) 
                         qstate nil))
        ))
\end{lstlisting}
Notice that Bob is completely unaware of the random numbers \texttt{r1} and
\texttt{r2}.  He knows only the result of measuring qubits $q_0$ and $q_1$.

The entire teleportation protocol is described in the following function:
\begin{lstlisting}
(defun quantum-teleportation-protocol 
       (alpha beta r1 r2)
  (let* ((qstate (quantum-teleportation-alice 
                   alpha beta r1 r2))
         (q1 (get-deterministic-qubit qstate 0))
         (q2 (get-deterministic-qubit qstate 1)))
    (quantum-teleportation-bob qstate q1 q2)))
\end{lstlisting}
The final verification of this protocol is that Bob ends up with a qubit ($q_2$)
in the same quantum state as Alice's initial qubit ($q_0$).  This is established
by the following theorem:
\begin{lstlisting}
(defthm quantum-teleportation-protocol-works
  (implies (and (acl2-numberp alpha)
                (acl2-numberp beta)
                (equal (+ (* alpha (conjugate alpha)) 
                          (* beta  (conjugate beta)))
                       1))
           (equal (narrow-to-qubit 
                   (quantum-teleportation-protocol 
                    alpha beta r1 r2) 
                   2)
                  (make-qubit alpha beta)))
  :hints ...)
\end{lstlisting}
The call to \texttt{narrow-to-qubit} extracts the state of qubit $q_2$
from the quantum state of all three qubits combined.  This is possible
only because Alice's antics broke the entanglement that was 
established earlier between qubits $q_1$ and $q_2$.

\section{Quantum Circuits: Evaluation}
\label{evaluator}

As we discussed in Sect.~\ref{teleportation}, the proof that the quantum teleportation
algorithm is correct is essentially one of symbolic evaluation.  That begs the question,
why can we not use direct evaluation?

The problem, of course, is that the computation involves the number $\sqrt{2}$, and possibly
other intermediate irrational numbers.  Nevertheless, having an evaluator that can at least
approximate quantum execution is valuable in itself, if only to debug circuits before attempting
a (laborious) verification effort\footnote{The authors would like to thank Dave Greve for
suggesting building an evaluator to complement this work.}.

We have experimented with various ways of implementing such an evaluator in ACL2.  One possibility
is to offer support for computation with the real numbers, e.g., by creating a representation for
irrational numbers and extending the arithmetic operators to work on this representation.  For
example, other theorem provers construct the reals (and other fields) by extending the rationals
either through adjoining a finite number of irrationals or introducing the classical construction
of the reals from analysis (cf.~\cite{Har:phd}).  This is a very intriguing possibility, and we have
experimented with this idea.  But it remains very much a topic for future work.

Instead, the approach we took was to use the existing function \texttt{(iter-sqrt x e)}, which calculates the square
root of \texttt{x} to an accuracy of \texttt{e} (for suitable arguments \texttt{x} and \texttt{e},
of course).  This function was previously defined using the bisection algorithm.  In fact, this
function is the basis for the definition of the square root function in ACL2(r)---essentially,
square root is defined by calling \texttt{(iter-sqrt x eps)} with an infinitesimal \texttt{eps}.

Then we mirrored the gate and normalization functions with an \texttt{-e} version, e.g.,
\texttt{qustate-norm-e}, \texttt{qu-H-gate-e}, and so on.  These versions accept an extra argument
\texttt{e} and replace calls to \texttt{acl2-sqrt} with corresponding calls to \texttt{iter-sqrt}.

The resulting system is cumbersome, but effective in evaluating ground instances of
quantum circuit, e.g., the teleportation algorithm with initial state $\alpha=\beta=\frac{131072}{185363}\approx\frac{1}{\sqrt{2}}$.
But the results are disappointing for a number of reasons.  First, the input and output uses fractions such as $\frac{131072}{185363}$
instead of the more attractive 0.70711.  Second, the connection between the evaluator and
the logical model is not complete.  Due to roundoff error, it's possible that the evaluator will
measure $\ket{0}$ when the model using the ``same'' inputs would measure the corresponding
qubit as $\ket{1}$.  This could happen in an intermediate computation, but resulting in a totally
different final quantum state.  Finally, all constant values are rational, and this includes the initial
state and any ``random'' numbers used for measurements.

\section{Unanswered Questions}
\label{conclusion}

We have described our ongoing formalization of quantum circuits in ACL2.  This includes a logical
model sufficient to reason about some quantum algorithms, such as the quantum teleportation
protocol.  It also includes a simulator that approximates the execution of quantum circuits.

In the previous section, we discussed some shortcomings of the current implementation, and
that leads naturally to possibilities for future work.  We are currently investigating several different
directions.  A natural direction is to formalize other quantum algorithms, e.g., Grover's search
algorithm and Shor's factoring algorithm.  It is worth noting that most of the proof of Shor's
algorithm is actually pure number theory, and it would be very  much in the style of previous
ACL2 formalizations.  Grover's algorithm, on the other hand, could be proved using the same
symbolic techniques used in this paper.

A second direction is to enhance ACL2(r) by providing support for computing with irrationals.  Our
first approach was to add needed irrationals one by one, e.g, $\sqrt{2}$, then $\sqrt{5}$, then
$\sqrt{\sqrt{2}+3}$, and so on.  The process would be one of extending an existing field by
adjoining each irrational, building a tower of fields, e.g., $\mathbb{Q} \subset \mathbb{Q}[\sqrt{2}]
\subset \mathbb{Q}[\sqrt{2},\sqrt{5}] \subset \cdots$.  Each step in the construction generates 
a field that can be represented using polynomials in the adjoined irrational, essentially terms
such as $3\sqrt{2}+5$.  However, there are some issues that crop up to ensure that the
representation is unique, e.g., if we mindlessly adjoin $\sqrt{2}$, $\sqrt{3}$, and $\sqrt{6}$.

These problems can be worked out, and it may even be possible to use some some of the
libraries developed for other theorem provers, e.g., \cite{AkPa:metitarski}.  It may be more
useful, however, to invoke a computer algebra system from ACL2(r) in order to simplify
such expressions.  Combinations of theorem provers and computer algebra systems have
already been done, e.g., \cite{Maple-PVS:TPHOLS01, KaFr:cas-hol}.  We believe such an
integration would be quite useful for ACL2(r) as well.

The third direction addresses the elephant in the room, so to write.  Quantum measurement
makes use of random numbers, and we have thus far completely ignored their genesis.
We are exploring the possibility of implementing some pseudo-random number generators
in ACL2.  Existing approaches to random number generation produce rational numbers.
In fact, they often produce integers, which are then scaled into the range $[0,1]$.  Implementing
a handful of pseudo-random number generators in ACL2 would not be difficult.  More challenging,
however, is deciding what can be said about such functions.  We intend to approach this by
reasoning about lists of random numbers, not the generators themselves.  Our hope is that we
will be able to prove certain properties of these lists, e.g., that according to a $\chi^2$ test,
the generated list is likely to have the underlying distribution $U[0,1]$.

Finally, we mentioned previously that linear algebra provides a unifying theme that can be
used to study quantum circuits.  In this setting, quantum states are vectors, gates are unitary
matrices, and measurements correspond to Hermitian matrices, Eigenvalues, and Eigenspaces.
In a separate collaboration, a colleague has been investigating rotation matrices, and it is possible
that much of that formalization can be reused in this context.  This unifying treatment may result in
much more elegant proofs of the correctness of quantum circuits.

\bibliographystyle{eptcs}
\bibliography{quantum} 

\end{document}